\documentclass[9pt,twocolumn,twoside]{pnas-new}
\usepackage{tikz,pgfplots}
\usetikzlibrary{external}
\usetikzlibrary{pgfplots.groupplots}
\usepgfplotslibrary{units}
\usepackage{filecontents}
\pgfplotsset{compat=1.5}

\newcommand{\highlight[2]}{\textcolor{black}{\textit{#2}}}

\templatetype{pnasresearcharticle} 

\title{Microemulsion sponge phase as a manifestation of the superflexibility critical point of tensionless balanced liquid-liquid interfaces}

\author[a,1]{Ramanathan Varadharajan}
\author[a]{Frans A. M. Leermakers} 

\affil[a]{Physical Chemistry \& Soft Matter, Wageningen University and Research Center, Stippeneng 4, 6708 WE Wageningen, The Netherlands.} 

\leadauthor{Varadharajan} 

\significancestatement{
In microemulsions, water and oil can mix up to the nanometer length scale and the interfaces may adopt a complex topology. We are so accustomed to the above idea that the key fundamental question, `what controls their behavior?', remains unanswered. We show, by model calculations, that the spectacular properties of microemulsion sponges are fundamentally linked to a critical point (CP). This CP is coupled to the (in)ability of carefully selected surfactants to generate, upon adsorption onto the oil-water interface, the tensionless state. As the interfacial rigidities vanish at the critical point , we refer to it as the superflexibility CP (SFP). Other systems that have `balanced' and `tensionless' interfaces, such as biomembranes, are expected to be influenced by similar SFP's.} 

\authorcontributions{F.A.M.L designed research, R.V performed calculations and analyzed data, R.V and F.A.M.L prepared the manuscript.}
\authordeclaration{Please declare any conflict of interest here.}
\correspondingauthor{\textsuperscript{1}To whom correspondence should be addressed. E-mail: ramanathan.varadharajan@wur.nl}

\keywords{microemulsion $|$ mean-field theory $|$ critical behavior $|$ sponge phase $|$ bending rigidities } 

\begin{abstract}
We have analyzed the mechanical properties of surfactant loaded tensionless balanced liquid/liquid interfaces using self-consistent field theory implementing a coarse grained model. Such a tensionless state, as occurring in microemulsions, signals a first-order interfacial phase transition. Consequently, the interfacial area is set by the amount of surfactant in the system. Near the bulk critical point, such systems suffer an interfacial, so-called superflexibility critical point (SFP) as soon as the tensionless state ceases to exist. This happens when the width of the L/L interface becomes comparable to the surfactant size. The bending rigidities vanish upon approaching the SFP as a power-law. Exclusively near the SFP, the saddle splay modulus is positive and therefore the sponge phase is its main characteristic.  
\end{abstract}

\dates{This manuscript was compiled on \today}
\doi{\url{www.pnas.org/cgi/doi/10.1073/pnas.XXXXXXXXXX}}

\begin{document}

\maketitle
\thispagestyle{firststyle}
\ifthenelse{\boolean{shortarticle}}{\ifthenelse{\boolean{singlecolumn}}{\abscontentformatted}{\abscontent}}{}

\dropcap{M}icroemulsions are thermodynamically stable systems of oil, water, and surfactants \cite{danielsson1981definition,yamamoto2001transparent,thiam2013biophysics}. They are often found either as droplets of oil in water or as water in oil, depending on the amount of their components. However, in the presence of comparable amount of oil and water and at high surfactant loading the interfacial tension between the oil and water becomes ultra-low, ideally it becomes tensionless. When in addition such a system is balanced by adding co-solvents, there is no significant tendency of the interfaces to curve in a preferred direction. Then a lamellar phase with a huge internal area is formed wherein oil and water layers alternate and the surfactant is predominantly at the (tensionless) interfaces between these domains \cite{safran1986origin}. It may happen that oil and water domains are fused due to the presence of handles and holes and the system is bicontinuous, we then refer to this as the microemulsion sponge phase. The middle phase of a Winsor III system often has such features \cite{scriven1976equilibrium, clausse1981bicontinuous, dave2007self, jones2012nanocasting}.

In a recent paper \cite{varadharajan2018sign} we have used the self-consistent field theory and implemented a united atom model for microemulsions featuring two complementary solvents $A_n$, $B_n$ and a symmetric copolymer $A_NB_N$. In this system there is just one Flory-Huggins interaction parameter $\chi \equiv \chi_{AB} \propto 1/T$ ($\chi>0$ implies repulsion). We have used significantly large values so that the solvents have a solubility gap and an interface develops between the A$_n$- and B$_n$-rich phases. Addition of surfactants leads to spontaneous adsorption at the interface and consequently the interfacial tension drops. The interfacial area remains microscopic as long as the tension is positive: $\gamma >0$. Typically in experiments when the balance is not perfect one may expect a droplet phase to form with increasing surfactant composition; this may occur in a narrow yet finite range of surfactant coverages/surfactant chemical potentials. However, when due to the symmetry such droplet phases are suppressed (in balanced systems), the tension vanishes at a well-defined amount of adsorbed surfactants/specified chemical potential of the surfactant. Then the system suffers a \textit{first order phase transition} as the area between oil and water grows jump-like to macroscopic values (it becomes simply proportional to the amount of surfactants in the system). Such extensive interface is characteristic for the microemulsion middle phase, the lamellar phase and the sponge phase.

Physics of such tensionless interfaces is dictated by their bending rigidities. Using the model of symmetric tensionless interfaces, we were able to unambiguously evaluate the bending rigidities of isolated interfaces in high precision \cite{varadharajan2018sign}. These moduli feature in a Taylor series expansion of the interfacial tension $\gamma$ in the mean $J = 1/R_1 + 1/R_2$ and Gaussian curvatures  $K=1/R_1\cdot 1/R_2$ known as the Helfrich equation (cf Eqn \ref{Helfrich} below). The mean bending rigidity $\kappa$ and Gaussian or saddle splay modulus $\bar{\kappa}$ control the shape fluctuations and the topology, respectively, of the interfaces. Consistent with expectations, the $\kappa$ was found to be positive. We reported a sign switch of $\bar{\kappa}$ from negative in the strong segregation case, i.e. $\chi \gg \chi^{\rm cr}$,  to positive when the solvents are weakly segregated, i.e. $\Delta \chi = \chi-\chi^{\rm cr}$ was reduced ($\chi^{\rm cr}$ is the  critical interaction parameter). The negative $\bar{\kappa}$ was suggested to be coupled to the lamellar microemulsion phase, while the positive value pointed to the stability of the sponge phase, or the bicontinuous microemulsion phases in general. 

It is of interest to know the interfacial rigidities of tensionless balanced interfaces when the system approaches the bulk critical point. It was decided to focus on the limit $N>n$ where the surfactants adsorb strongly at the interfaces. When this systematic work was underway, \highlight{it was found that it is only possible to reach the tensionless state for sufficiently high values of} $\chi \ge \chi^{\rm s}>\chi^{\rm cr}$. We realized that the lower limit $\chi=\chi^{\rm s}$ is a special point for microemulsions. We argued above that the value of $\gamma=0$ is a point of a first-order phase transition. When $\chi $ (temperature) is our control parameter, we can construct a phase diagram, e.g. by plotting the adsorbed amount of surfactant to reach the tensionless state as a function of the tuning parameter $\chi$. In this phase diagram a line of first order phase transitions stops at a critical point $\chi = \chi^{\rm s}$, which in itself is a \textit{second order phase transition}. In this work, we thus establish such a generic critical point. Importantly, we show scaling behavior of the rigidities towards this critical point and therefore coin this critical point as superflexibility  critical point, SFP, also to avoid confusion with the bulk critical point. We will argue below that the sponge phase is a manifestation of the SF-criticality.

Predicting the value of the bending rigidities is the first step, to have models that uses these quantities to rationalize the microemulsion phase diagram is the second step. To model, e.g., bicontinuous phases of microemulsions, a random description of interface was used in earlier theoretical works \cite{talmon1978statistical,de1982microemulsions}. Talmon-Prager model divides the spatial-volume using Voronoi tessellation and creates random interface by filling the space with either oil or water \cite{talmon1978statistical}. Later this model was improved to explicitly account for amphiphilic surfactants and interfacial curvature energy. Though several theoretical works report to have successfully modeled middle-phase/sponge phase microemulsions, the rigidities provided as inputs in these models deviate from our current and recent predictions \cite{varadharajan2018sign}. Hence we argue that these models are not yet optimally exploited. Again, as the  genuine rigidities couldn't be provided, the models could not anticipate the existence of an SFP. We trust that when the SFP scenario as elaborated below will be implemented, we might get a more complete picture of microemulsions. 

Several experimental works are available targeted to determine the critical behavior of microemulsion systems. Dynamic light scattering experiment proved that the phase transition near the cloud-point temperature features characteristics of a critical phenomenon \cite{huang1981critical}, and critical end points for a microemulsion phase in equilibrium with aqueous phase and organic phase were reported \cite{cazabat1982critical}. As the interfacial rigidities vanish at the SFP, we do expect \textit{some} characteristic scattering signature in the vicinity of the SFP. In line with this it is known that the sponge phase is weakly turbid \cite{strey1990dilute,clausse1981bicontinuous}. Nonetheless our results suggest that near the SFP, microemulsion system will \highlight{not} feature classical critical behavior. 

We note that in the absence of the copolymer, our model reduces to a Van der Waals like system (with critical point $\chi^{\textrm{cr}}=2/n$ and a critical volume fraction $\varphi^{\rm cr}=0.5$) which allows analytical evaluation of its interfacial properties near the bulk critical point. It is well known that the interfacial tension vanishes as a power-law with (mean field) coefficient $3/2$, i.e., $\gamma \propto \Delta \chi^{3/2}$. The density difference between the two phases vanishes with a (mean field) coefficient $1/2$, while the width of the interface diverges as the mean field coefficient is $-1/2$ \cite{szleifer1988curvature,blokhuis1993van,gompper1992ginzburg,gompper1995self}. In addition, both analytical theory and numerical SCF \cite{oversteegen2000rigidity} revealed that $-\kappa$ and $\bar{\kappa }$ vanish with (van der Waals) coefficient ${1/2}$. Again these coefficients are the mean field values. The ratio $|\kappa/\bar{\kappa}| = (\pi^2-3)/(\pi^2-6)\approx 2$, See section 4 of supplemental material \cite{supplemental}. Obviously, this bulk critical behavior plays an important role in what follows and we will refer to this van der Waals scenario regularly.


\section*{Results and Discussion}
Our results are generated using the self-consistent field theory of Scheutjens and Fleer (SCF-SF) and we employed the same molecular model (mentioned above) as in our previous work. The focus is therefore not on the methodology but rather to the critical behavior upon approach towards the newly identified SFP. See \textit{Materials and methods} for details about computational methods and model.
\begin{figure}[t]
\begin{tikzpicture}
\begin{axis}[
  x unit={},
  y unit={},
  xlabel={z},
  ylabel={x,y}, 
  y label style={at={(-0.275,0.5)},anchor=south},
  yticklabels={,,},
  legend pos=north west,
  minor x tick num=1,
  minor y tick num=1, xmin=-30, xmax=30, ymin=-30, ymax=30,width=0.5\columnwidth] 
\shade[left color=red,right color=yellow, opacity=0.75] (axis cs: -30,-30) rectangle (axis cs: 0,30);
\shade[left color=yellow,right color=blue, opacity=0.75] (axis cs: 0,-30) rectangle (axis cs: 30,30);
\node at (axis cs: 30, 0.65) [anchor=north east] {(a)};
\draw[|<->|,black,thick] (axis cs:5,0.7) -- node[above]{W$<$N} (axis cs:-5,0.7);
\end{axis}
\end{tikzpicture}
\begin{tikzpicture}
\begin{axis}[
  x unit={},
  y unit={},
  xlabel={z},
  ylabel={x,y}, 
  y label style={at={(-0.275,0.5)},anchor=south},
  yticklabels={,,},
  legend pos=north west,
  minor x tick num=1,
  minor y tick num=1, xmin=-30, xmax=30, ymin=-30, ymax=30,width=0.5\columnwidth] 
\shade[left color=orange,right color=yellow, opacity=0.75] (axis cs: -30,-30) rectangle (axis cs: 0,30);
\shade[left color=yellow,right color=cyan, opacity=0.75] (axis cs: 0,-30) rectangle (axis cs: 30,30);
\node at (axis cs: 30, 0.65) [anchor=north east] {(d)};
\draw[|<->|,black,thick] (axis cs:10,0.7) -- node[above]{W$\approx$N} (axis cs:-10,0.7);
\end{axis}
\end{tikzpicture}

\begin{tikzpicture}
\begin{axis}[
  x unit={},
  y unit={},
  xlabel={z},
  ylabel={$\varphi$}, 
  y label style={at={(-0.275,0.5)},anchor=south},
  legend pos=north west,
  minor x tick num=1,
  minor y tick num=1, xmin=-30, xmax=30, ymin=0, ymax=1, width=0.5\columnwidth]
\fill [fill=yellow, opacity=0.5] (axis cs: 5,0) rectangle (axis cs: -5,1);
\addplot[color=red, thick] table [x index=0,y index=1] {phi_fc.dat};
\addplot[color=blue, thick] table [x index=0,y index=2] {phi_fc.dat};
\addplot[color=green, thick] table [x index=0, y index=3] {phi_fc.dat};
\node at (axis cs: 30, 0.65) [anchor=north east] {(b)};
\end{axis}
\end{tikzpicture}
\begin{tikzpicture}
\begin{axis}[
  x unit={},
  y unit={},
  xlabel={z},
  ylabel={$\varphi$}, 
  y label style={at={(-0.275,0.5)},anchor=south},
  legend pos=north west,
  minor x tick num=1,
  minor y tick num=1, xmin=-30, xmax=30, ymin=0, ymax=1,width=0.5\columnwidth]
\fill [fill=yellow, opacity=0.5] (axis cs: 15,0) rectangle (axis cs: -15,1);
\addplot[color=red, thick] table [x index=0,y index=1] {phi_nc.dat};
\addplot[color=blue, thick] table [x index=0,y index=2] {phi_nc.dat};
\addplot[color=green, thick] table [x index=0, y index=3] {phi_nc.dat};
\node at (axis cs: 30, 0.65) [anchor=north east] {(e)};
\end{axis}
\end{tikzpicture} 

\begin{tikzpicture}
\begin{axis}[
  x unit={},
  y unit={},
  xlabel={z},
  ylabel={$\omega$}, 
  y label style={at={(-0.295,0.5)},anchor=south},
  legend pos=north west,
  minor x tick num=1,
  minor y tick num=1, xmin=-30, xmax=30,width=0.5\columnwidth]
\fill [yellow, opacity=0.5] (axis cs: 5,-0.004) rectangle (axis cs: -5,0.007);
\addplot[color=black, thick] table [x index=0,y index=1] {pressure_fc.dat};
\node at (axis cs: 30, 0.002) [anchor=north east] {(c)};
\end{axis}
\end{tikzpicture}
\begin{tikzpicture}
\begin{axis}[
  x unit={},
  y unit={},
  xlabel={z},
  ylabel={$\omega$}, 
  y label style={at={(-0.295,0.5)},anchor=south},
  legend pos=north west,
  minor x tick num=1,
  minor y tick num=1, xmin=-30, xmax=30,width=0.5\columnwidth]
\fill [yellow, opacity=0.5] (axis cs: 15,-0.00005) rectangle (axis cs: -15,0.00005);
\addplot[color=black, thick] table [x index=0,y index=1] {pressure_nc.dat};
\node at (axis cs: 30, -0.000001) [anchor=north east] {(f)};
\end{axis}
\end{tikzpicture} 

 \caption{Interfacial behavior.  (a,d) Graphical illustration of the density profiles across the tensionless interface with emphasis on the width $W$ of the interface and the volume fraction (density) gradients by color change. (b,e) Selected examples for the volume fraction $\varphi$-profiles (SCF result) of bulk phases (red and blue lines) and surfactant (green line) for balanced tensionless interfaces. (c,f) corresponding  grand potential density $\omega$ (minus lateral pressure) profiles across the interface. (a, b, c) far from critical $\chi = 0.6$, (d, e, f) close to critical $\chi = 0.52$. $n=4$ and $N=20$. The interfacial zone is indicated by a yellow band (b, c \& e, f).
 }
 \label{f1}
\end{figure}
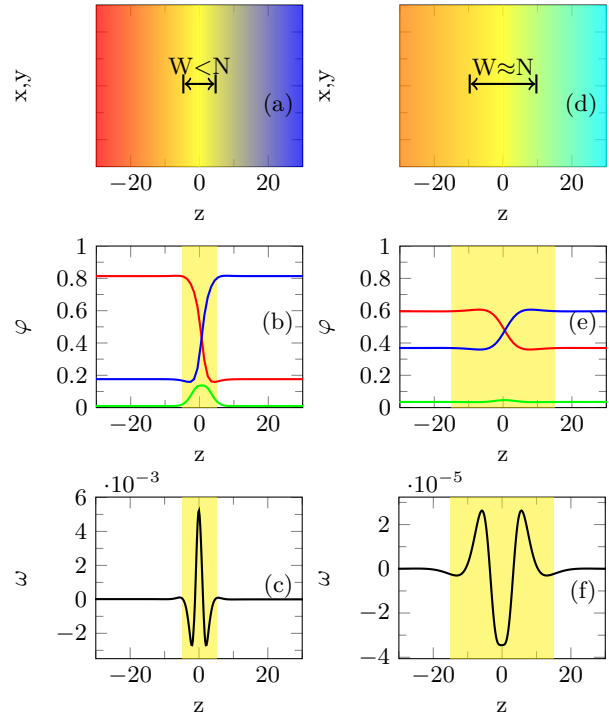
Let us first consider the planar A$_n$/B$_n$ interface and discuss what happens when we add the symmetric surfactants to this system. There is one interaction parameter $\chi$, which can be interpreted as an inverse temperature scale. With decreasing $\chi$ (increased temperature) we go towards the bulk critical point. The two phases gradually mixes and the width of the interface grows. Schematically this is illustrated in Fig. \ref{f1} (a, d). Surfactants that are added to the system spontaneously adsorb onto the interface and consequently the interfacial tension decreases. The (for microemulsions) relevant equilibrium point is characterized by the tensionless state. In fig. \ref{f1}b,e we give the density distributions as found by the SCF formalism for such equilibrium point for a system far from critical and much closer to critical, respectively. It is noticed that significantly more surfactants are needed in the strong segregation case than in the system that is close to critical to reach such tensionless state. From the density profile we can also see that the interface widens as illustrated in viewgraphs (a,d), and the density difference between the two phases diminishes. In fig. \ref{f1}c,f we report the corresponding grand potential density profiles $\omega (z)$. As the tension is zero, the $\omega$-profiles should have positive and negative `excursions'. Moreover, the profiles are symmetric with respect of the $z_0$-plane, i.e. $\omega (z-z_0) = \omega (z_0-z)$. The Gaussian bending rigidity is the second moment of this grand potential density profile and is non-zero for both cases. In fact $\bar{\kappa}$ has a negative value for the strong segregation and a positive value for the weak segregation case. The interesting evolution of the $\omega(z)$ profiles as a function of $\chi$ is given in supplementary information and is not the topic of the present paper.
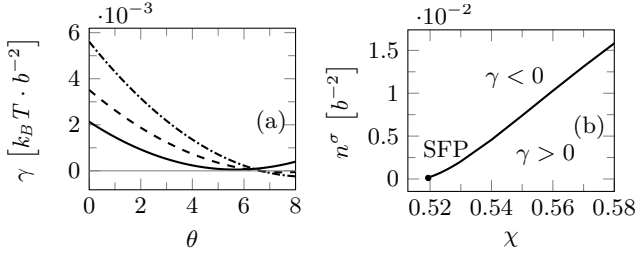
\begin{figure}[t]
\centering
\begin{tikzpicture}
\begin{axis}[
  x unit={},
  y unit={{\it k_BT\cdot b}^{-2}},
  xlabel={$\theta$},
  ylabel={$\gamma$}, 
  y label style={at={(-0.2,0.5)},anchor=south},
  legend pos=north west,
  minor x tick num=1,
  minor y tick num=1, xmin=0, xmax=8, width=0.5\columnwidth]
\addplot[color=black, thick] table [x index=0,y index=2] {superflexibility.dat};
\addplot[color=black, thick, dashed] table [x index=0,y index=3] {superflexibility.dat};
\addplot[color=black, thick, densely dashdotted] table [x index=0, y index=4] {superflexibility.dat};
\addplot[color=gray, thin] table [x index=0, y index=1] {superflexibility.dat};
\node at (axis cs: 8, 0.003) [anchor=north east] {(a)};
\end{axis}
\end{tikzpicture}
\begin{tikzpicture}
\begin{axis}[
  x unit={},
  y unit={{\it b}^{-2}},
  xlabel={$\chi$},
  ylabel={$n^\sigma$}, 
  y label style={at={(-0.2,0.5)},anchor=south},
  legend pos=north west, xmax=0.58,
  minor x tick num=1,
  minor y tick num=1, width=0.5\columnwidth]
\addplot[color=black, thick] table {adsamountvschi.dat};
\draw plot [mark=*,cyan!50!black, mark size=1, opacity=0.3] coordinates{(0.518875,0.00001178856172250)};
\node at (axis cs:0.514875,0.0055) [anchor=north west] {SFP};
\node at (axis cs:0.56,0.014) [anchor=north east] {$\gamma < 0$};
\node at (axis cs:0.57,0.005) [anchor=north east] {$\gamma > 0$};
\node at (axis cs: 0.58, 0.0085) [anchor=north east] {(b)};
\end{axis}
\end{tikzpicture}
 \caption{Thermodynamical properties near SFP. (a) Interfacial tension, $\gamma$ [in units of $k_BT/b^2$ with $b$ a segment length] of the planar interface as a function of the amount of surfactant segments per unit area $\theta$ in the system. $n=4$, $N=20$. Near critical interaction parameters are: $\chi= 0.52$ (solid), $0.53$  (dashed) $0.54$ (dash-dot).
(b) Phase diagram in (excess) adsorbed number of surfactants (per unit area) in the tensionless state $n^{\sigma}$ - interaction parameter $\chi$ coordinates. A line of first-order phase transitions, i.e. ($n^{\sigma}$, $\chi$) values that lead to the tensionless state $\gamma=0$, terminates at a critical point (SFP) indicated by a black dot. Above the line the interfacial tension is negative, below the line it is positive.
}
 \label{f2}
\end{figure}
Let's next discuss more precisely what happens to the interfacial tension when surfactants A$_{20}$B$_{20}$ are added to the A$_4$/B$_4$ interface, and focus on the \highlight{critical region}. The system size is chosen large enough so that the surfactants at the (planar) interface do not `see' the system boundaries. We fix the total volume in the system and then add the surfactants (and thus reduce the amounts of the two bulk phases to accommodate for the surfactant). The amount of surfactant segments in the system (per unit interfacial area) obeys to $\theta = \int \varphi_s(z) {\rm d}z$ with $\varphi_s(z)$ the surfactant density profiles as found by the SCF formalism. This amount splits up into an adsorbed amount of surfactant segments per unit area $\theta^\sigma = \int (\varphi_s(z)-\varphi_s^b) {\rm d}z$ and an amount per unit area in the bulk phases $\theta^b\equiv \theta-\theta^\sigma$. Note that by symmetry the polymer concentration in the A-rich bulk equals that in the B-rich bulk and there is need to identify a Gibbs plane, i.e. the excess is (by symmetry) already perfectly uniquely defined. 

In Fig. \ref{f2}a we present selected results wherein the interfacial tension is plotted as a function of the amount of surfactant in the system. Typically a decreasing trend is found, for which the tension is positive at low $\theta$ and becomes negative at high values (e.g. for $\chi=0.54$). When the interaction parameter is reduced, the interfacial tension of the bare L/L interface decreases and fewer polymers are needed to find the tensionless state (consistent with the result of Fig. \ref{f1}). This trend does not continue. As can be seen in Fig. \ref{f2}a the tension in fact goes through a local minimum. This minimum is possible as with increasing $\theta$ the value of $\theta^b$ increases rather than $\theta^\sigma$; in fact $\theta^\sigma$ may go through a lowest value itself, because the bulk concentration increases (e.g. for $\chi=0.52$). It thus is feasible that the minimum value in $\gamma(\theta)$ touches the $\gamma=0$ value. This happens at a special interaction parameter $\chi^{\rm s}$. When $\chi<\chi^{\rm s}$ the interfacial tension first decreases but then increases with increasing amount $\theta$, while for all cases $\gamma >0$. This happens for all super-critical values for $\chi$, i.e. for $\chi < \chi^{\rm s}$.

In fig. \ref{f2}b we present for specified interaction parameter $\chi$ ($x$-axis) the number of surfactants per unit area $n^\sigma = \theta^\sigma /2N$ that are needed to find the tensionless state ($y$-axis). This viewgraph serves as a phase diagram wherein the line represent a line of tensionless states. We realize that the interfacial tension is the first derivative of the free energy with respect to the area of the interface. \highlight{When this first-order derivative vanishes, it represents a first-order phase transition.} The line in Fig. \ref{f2}b thus represents a line of first-order phase transitions. It runs from a high surface coverage at large values of $\chi $ to extremely low surface coverage at lower $\chi$. More exactly, the line ends at the critical point, which for $n=4$, $N=20$, is at $\chi^* \approx 0.518875$ and $n^\sigma \approx 1.1788 \times 10^{-4}$ (which is significantly above zero). Such critical point is largely unexplored and turns out to be of central importance for microemulsion systems. Before we deal with the many consequences we want to elaborate why the SFP occurs in the first place.
 
\begin{figure}[t]
 \centering
\begin{tikzpicture}
\begin{loglogaxis}[
  x unit={},
  y unit={},
  xlabel={$N$},
  ylabel={$\Delta \chi^{{\rm s}-{\rm cr}}$}, 
  xmax=100,
  xmin=2,
  legend pos=north east,
  y label style={at={(-0.2,0.7)},anchor=south},
  minor x tick num=1,
  minor y tick num=1, width=.5\columnwidth]
\addplot[color=black, thick] table [x index=0,y index=1] {sfpvsn.dat};
\addplot[color=black, thick, dashed] table [x index=0,y index=2] {sfpvsn.dat};
\addplot[color=black, thick, densely dashdotted] table [x index=0,y index=3] {sfpvsn.dat};
\draw[gray,thick] (axis cs:6,1e-1) -- node[black,left]{$\frac{dy}{dx}=-2$} (axis cs:60,1e-3);
\node at (axis cs: 1e2, 1) [anchor=north east] {(a)};
\end{loglogaxis}
\end{tikzpicture}
\begin{tikzpicture}
\begin{loglogaxis}[
  x unit={},
  y unit={{\it k_BT\cdot b}^{-2}},
  xlabel={$\Delta\chi^s$},
  ylabel={$k_A$}, 
  xmax=1e-2,
  xmin=1e-6,
  legend pos=north west,
  minor x tick num=1,
  minor y tick num=1, width=0.5\columnwidth]
\addplot[color=black, thick] table [x index=0,y index=1] {ka.dat};
\draw[gray,thick] (axis cs:6e-3,4e-4) -- node[black,left]{$\frac{dy}{dx}=2$} (axis cs:2e-3,4e-5);
\node at (axis cs: 2e-5, 2.5e-4) [anchor=north east] {(b)};
\end{loglogaxis}
\end{tikzpicture}
 \caption{Behavior of SFP and area expansion modulus. a) The difference of the SFP and the bulk critical point, $\Delta \chi^{{\rm s} - {\rm cr}}$ as a function of the surfactant block length, $N$ in double logarithmic coordinates. Lines are for: $n = 2$ (solid), $4$ (dashed), $8$ (dash-dotted). b) The area expansion modulus $K_A$ as a function of $\Delta \chi^{\rm s}$ in log-log coordinates for $n=4$ and $N=20$. The slopes are indicated by grey lines.}
 \label{f3}
\end{figure}
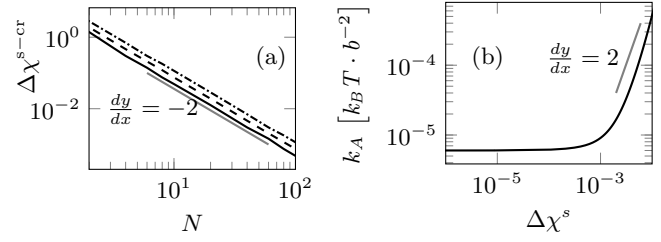

It is anticipated that the closeness of the SFP to the bulk critical point is a function of the block length of the surfactant. Motivated by this we present in Fig. \ref{f3}, the difference of the SFP and the bulk critical point, $\Delta \chi^{{\rm s}-{\rm cr}} \equiv \chi^{\rm s}-\chi^{\rm cr}=\chi^{\rm s}-2/n$, as a function of the length $N$ of the block of the surfactant in log-log coordinates. Results for $n=2,4,8$ are shown. For all the solvent sizes we find that $\Delta \chi^{\rm s} \propto N^{-2}$ to a good approximation. We know from van der Waals theory (see above) that the width $W$ of the A$_n$/B$_n$ interface is given by $W\propto (\chi-\chi^{\rm cr})^{-1/2}$ and that the numerical coefficient is a weak function of $n$. Inserting $\chi=\chi^{\rm s}$, we have $\Delta \chi^{{\rm s}-{\rm cr}} =W^{-2}$. Clearly when $W \propto N$, we arrive at $\Delta \chi^{{\rm s}-{\rm cr}} \propto N^{-2}$, which is the result of Fig. \ref{f3}. The numerical coefficient is again a small function of $n$, which can be traced to the for the width of the bare interface. We conclude that SFP is indeed triggered by the condition that the width of the interface is proportional to the surfactant size.

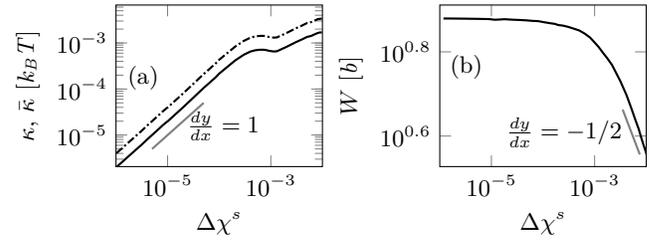
\begin{figure}[t]
\centering
\begin{tikzpicture}
\begin{loglogaxis}[
  x unit={},
  y unit={\it k_BT},
  xlabel={$\Delta\chi^s$},
  ylabel={$\kappa$, $\bar{\kappa}$}, 
  xmax=1e-2,
  xmin=1e-6,
  legend pos=north west,
  minor x tick num=1,
  minor y tick num=1, width=0.5\columnwidth]
\addplot[color=black, thick] table [x index=0,y index=1] {scaling.dat};
\addplot[color=black, thick, densely dashdotted] table [x index=0,y index=2] {scaling.dat};
\draw[gray,thick] (axis cs:5e-5,5e-5) -- node[black,right]{$\frac{dy}{dx}=1$} (axis cs:5e-6,5e-6);
\node at (axis cs: 1e-5, 5e-4) [anchor=north east] {(a)};
\end{loglogaxis}
\end{tikzpicture}
\begin{tikzpicture}
\begin{loglogaxis}[
  x unit={},
  y unit={\it b},
  xlabel={$\Delta\chi^s$},
  ylabel={$W$}, 
  xmax= 1e-2,
  xmin=1e-6,
  legend pos=north west,
  minor x tick num=1,
  minor y tick num=1, width=.5\columnwidth]
\addplot[color=black, thick] table [x index=0,y index=1] {iw.dat};
\draw[gray,thick] (axis cs:3.5e-3,4.6) -- node[black,left]{$\frac{dy}{dx}=-1/2$} (axis cs:7.5e-3,3.6);
\node at (axis cs: 1e-5, 6.5) [anchor=north east] {(b)};
\end{loglogaxis}
\end{tikzpicture}
 \caption{Critical scaling near SFP. a) The mean $\kappa$ (dash dotted) and Gaussian bending modulus $\bar{\kappa}$ (solid) in unit of $k_BT$, b) the width $W$ of the interface in units $b$, as a function of $\chi  - \chi^{\rm s}$ in log-log coordinates. (Here $n=4$ and $N=20$ for which $\chi^{\rm s}=0.518875$). The width is found from the planar interface as $W = |( \varphi^b_{A_n}-\varphi^b_{B_n} )/(\varphi_{A_n}(z_0-1/2)-\varphi_{A_n}(z_0+1/2))|$. Relevant slopes are indicated.  Slightly outside the plotting range, namely at $\chi \approx 0.5415$, the $\bar{\kappa}$ switches sign; the mean bending rigidity $\kappa$ remains positive for all $\chi$-values.}
 \label{f4}
\end{figure}

For microemulsions that have zero-tension interfaces, the bending rigidities are of central importance. In fig. \ref{f4} we present results for both moduli as a function of $\Delta \chi^{\rm s} \equiv \chi -\chi^{\rm s}$ for the near critical region where both $\bar{\kappa}$ and $\kappa$ are positive (fig. \ref{f4}a), in combination with the width of the interface $W$ (fig. \ref{f4}b). Going from high to low $\chi$, as predicted by the van der Waals theory, in the region $10^{-3}< \Delta \chi^{\rm s} <10^{-2}$, the interfacial width increases $W\propto (\Delta \chi^{\rm s})^{-1/2}\approx (\Delta \chi)^{-1/2}$ (because in this case $\chi^{\rm s}-\chi^{\rm cr}$ is small), and, importantly, both $\bar{\kappa}$ as well as $\kappa$ show scaling dependences $(\Delta \chi^{\rm s})^{1/2}$ (scaling exponent identical to the van der Waals interface). Interestingly, below  $\Delta \chi^{\rm s} < 10^{-3}$ the width of the interface saturates (in these coordinates) at a value for $W$ that is comparable to the length of the surfactant (see fig. \ref{f3}).  The moduli go through a local minimum and a weak maximum before they continue to decrease as a power-law with a new coefficient of unity: $\bar{\kappa} \sim \kappa \propto (\chi-\chi_{\rm s})^1$.

Importantly, at $\chi = \chi^{\rm s}$ both bending rigidities vanish. Interfaces that lack any rigidity are super-flexible and this is the reason why we have named this interfacial critical point the superflexibility critical point. Consistently, the ratio $\kappa/\bar{\kappa}\approx 2$ (a similarly number is found for the van der Waals L/L interface for the ratio of the absolute values of the rigidities). This ratio changes dramatically (not shown) when $\bar{\kappa}$ switches sign at higher $\chi$ than used in fig. \ref{f4}. We do not consider this here because the sign switch of $\bar{\kappa}$ was the topic of our previous paper \cite{varadharajan2018sign}. 

The interesting result that there are two power-law regions for a given parameter can be rationalized from the imposed constraint. From van der Waals theory we know that both $\kappa$ as well as $\bar{\kappa}$ show critical scaling upon approach to the bulk critical point with power-law coefficient $1/2$. Here we recover the same slope as long as the width of the interface can grow 
and we are in the `van der Waals' regime: the system is under control of the bulk critical point. However, as long as we insist on having tensionless interfaces, the width of the interface cannot grow without bounds. As soon as this constraint is felt by the system, it becomes under control of the SFP. The density difference between the two bulk values, the number of surfactants at the interface, etc., all become virtually constant and a new scaling region sets in this regime with a new power-law exponents of $1$ instead of $1/2$: the new power-law exponent manifests as a consequence of the imposed length scale. 

From the above we see that the SFP has non-classical features. To add to this we present results for the area expansion modulus $K_A \equiv \frac{\partial \gamma }{\partial \log A_s}= -\frac{\partial \gamma }{\partial \log \theta^\sigma }$ in Fig. \ref{f3}b as a function of $\Delta \chi^{\rm s}$ in double logarithmic coordinates. $K_A$ shows scaling behavior (as expected) in the van der Waals regime, that is for $\Delta \chi^{\rm s}>10^{-3}$ where a (mean field) slope of $2$ is found, but less trivially not in the region near the SFP. Even though $\gamma(\theta)$ goes through a minimum at the SFP (cf. fig. \ref{f2}a), $\gamma(n^{\sigma})$ has a cusp shape, while $n^{\sigma}$ is finite at the SFP (cf fig. \ref{f2}b). Hence, at the SFP the interfacial area fluctuations will not diverge.  

\subsection*{Implications on microemulsion phase behavior}
At the turn of the century Safran\cite{safran1999curvature} argued that the exchange of surfactant with the bulk will soften the interfaces and mentioned that this effect is important for weak adsorption. Besides this qualitative softening effect, we do not know any quantitative numerical estimates for bending rigidities in the weakly adsorbing situation. Invariably the rigidities are analyzed in the sharp interface approximation \cite{safran1999curvature, lekkerkerker1989contribution, szleifer1988curvature, birshtein2008curvature}. Near the SFP, however, the width of the interface is of the same order as the surfactant size and then the sharp interface approximation is not expected to be accurate. This might explain why the scaling behavior of the rigidities \textit{remained unnoticed} to date.

The idea that the tensionless state of the balanced L/L interface is of key importance to microemulsions is widely recognized. But more surprisingly the idea that the loss of the tensionless state represents a true critical point with accompanied scaling behavior for the rigidities remained largely unexplored. In the context of weak adsorption of surfactant at the oil-water interface Safran discussed \cite{safran1999curvature} a critical point using an analogy with the critical volume fraction of surfactants above which micelles form, in this case, above which interfaces can develop. To us this analogy is ambiguous because this might apply to all points on the line in our phase diagram (cf. Fig. \ref{f2}b) and not only to $\chi=\chi_s$. Safran argued that the interfacial rigidities have a finite value only above this critical surfactant concentration \cite{safran1999curvature}. This result matches our prediction only at the SFP, but it is in conflict with it for all other values of the interaction parameter. Indeed we argue that the control parameter is \textit{not} the surfactant concentration in the system, but rather an `effective' strength of adsorption/interaction, embodied in $\chi$, that is the natural control parameter to reach microemulsion criticality. Nor the idea that an SFP condition naturally develops near the bulk critical point, nor the role of the surfactant length that determines $\Delta \chi^s$ (a result that is linked to our simple model with only one interaction parameter), nor the critical scaling of the rigidities, nor the (generic) sign switch of $\bar{\kappa}(\chi)$ at stronger segregation, were elaborated before and therefore the impact of an SFP for microemulsions is not appreciated yet. 

Typically in microemulsion systems very short surfactants are used to generate bicontinuous emulsions \cite{scriven1976equilibrium,gompper1995self}. From the above it is clear why this is a good idea. For such a system the SFP occurs far from the bulk critical point (cf fig.\ref{f3}) and hence one can easily observe microemulsion sponge phases. In practice this means that it is not necessary to add huge amounts of co-solvents to reduce the difference in polarity between the two bulk phases (or increase $T$) and still come in the critical zone where a sponge phase is possible. This trend was noticed before \cite{scriven1976equilibrium,gompper1995self}, however, we feel that our interpretation is more physical than the rational presented earlier. 

Irrespective of the surfactant size, we find the following (generic) scenario for the rigidities: (i) Far from the bulk critical point and also far from the SFP, $\kappa$ is high and positive and $\bar{\kappa}$ is negative (result highlighted in our previous paper). As argued earlier this implies that the lamellar phase is the expected topology. Repulsive undulation forces (increasing with decreasing $\kappa$) versus attractive Van der Waals forces will set the capabilities of this lamellar phase to swell (pick up water and/or oil). (ii) Upon approach to the bulk critical point (this can be done by adding co-solvents, or by increasing the temperature -implying a decrease in $\chi$-), the value of $\kappa $ decreases, but remains positive. The lamellar phase is expected to swell due to the increased Helfrich repulsion. At some point the $\bar{\kappa}$ switches sign and becomes positive. From hereon we might expect bicontinuous phases to occur.  
(iii) Upon approach towards the SFP first $\bar{\kappa}$ becomes of order $\kappa$ and after that both values decrease towards zero. Hence, the interfaces become highly flexible while the tendency to form handles and holes exist (but tend to decrease). A positive $\bar{\kappa}$ is expected for a sponge phase. Hence, sponge like microemulsions are expected to exist up to the SFP. Near the SFP the interfaces should strongly fluctuate and the characteristic distance between the handles should be related to the interface persistence length $l_p$. The latter decreases with decreasing value of $\kappa$ ($l_p \propto \exp \kappa$), and therefore the sponge phase is expected to occur at higher surfactant concentrations when the SFP is approached. 

These theoretical trends correlate extremely well with known experimental observations. It is instructive to mention a few of these. It is well known that for weak surfactants (short non-ionics) the one-phase  microemulsion system is typically bicontinuous, that is, sponge-like \cite{gompper1995self}. This result is understood from the sign switch of $\bar{\kappa}$ that occurs for short surfactants for $\chi$ far from critical, as discussed in our previous paper \cite{varadharajan2018sign}. Upon approach to the critical point the microemulsion phase shifts to higher surfactant concentrations \cite{kahlweit1985phase, kahlweit1988general, kahlweit1988properties}. As mentioned already this result is caused by a decrease of the bending rigidity upon approach to the SFP. For stronger surfactants (longer non-ionics), the lamellar phase $L_{\alpha}$ gradually becomes more pronounced. $L_{\alpha}$ first appears at low temperatures and high surfactant concentrations. Again this result is explained by the sign switch of $\bar{\kappa}$ discussed in our previous paper \cite{varadharajan2018sign}. At some higher temperature the lamellar phase swells and takes up most of the microemulsion phase volume, leaving only room for the sponge phase at rather thin slap in the microemulsion phase volume, namely at relatively low surfactant concentrations and relatively high temperatures \cite{kahlweit1986search}. This result is understood from the scaling dependences of the rigidities. The larger (stronger) surfactants have (for given $T$) lower (or more negative) values for $\bar{\kappa}$ \cite{varadharajan2018sign}, while shorter (weaker) surfactants are, for given $T$, closer to the SFP and therefore are more likely to support the sponge phase. 

Our current theoretical predictions are restricted to the balanced systems that by construction do not have a preferred curvature, but extensions to more realistic system can be envisioned. In reality such balanced systems do not occur and we have to carefully tune the amount of cosolvents to guide the system in the direction of zero spontaneous curvature. That is why in experiments there is not a simple variable as the temperature to control the microemulsion phase behavior \cite{strey1996phase}. Nevertheless our toy model wherein the balanced system is imposed is instructive. We can straightforwardly go towards the SFP and by doing so prove that the SFP is in charge of the microemulsion phase behavior, that is, the interface rigidity and topology. Importantly for all systems that we have considered we see the same phenomenology. Near the critical point the sponge like topology dominating the microemulsion structure. Far from the critical point we envision a lamellar topology. The region near the critical point can in general be approached with multiple control parameter. The temperature, the amount of cosolvent, etcetera. The sponge dome in the vicinity can be reached in several ways. 

Near the SFP both $\kappa$ as well as $\bar{\kappa}$ depend linearly on $\Delta \chi^{\rm s}$. It is understood that the scaling exponent (unity) is the mean field prediction. However, as the width of the interface does not diverge for the SFP, the intrinsic rigidities $\kappa$ and $\bar{\kappa}$ may well be relatively accurately predicted by SCF.  We know that in reality the effective rigidities are length scale dependent \cite{helfrich1985persistencelength, helfrich1987persistencelength} and typically interfaces appear more flexible on length scales larger than their persistence length. Interestingly however, as the (intrinsic) rigidities vanish at the SFP, the importance for renormalisation of the rigidities on length scales larger than the persistence length appears less of an issue for interfaces near the SFP. We trust that the current predictions for the interfacial rigidities will be instrumental to develop more quantitative models \cite{cates1988random,talmon1978statistical}.

Microemulsions have a spectacular phase behavior. Because we know this already for many years, we've got used to the extraordinary properties of these systems. We can mix oil and water (very complementary species) upto the nanometer scale and still `walk' through the system in any direction without leaving the oil nor the water phase (that is what the sponge phase allows us to do). The very close proximity of water and oil phases gives us the opportunity to make use of the complementary solvency conditions. There are already many applications of microemulsion that effectively make use of these amazing properties. This paper only offers an explanation of why this happens. 

Balanced tensionless interfaces occur also in other relevant systems. Importantly freely dispersed lipid bilayer membranes, and the biological counterparts, obey to the same characteristics. It is therefore foreseen that these structures also feature an SFP. In line with this, we know that special surfactant systems, such as C$_{12}$E$_5$, have a sponge phase very similar to the sponge phase in microemulsions \cite{strey1990dilute}. Indeed bilayer sponges are expected for the very same reason as these occur in microemulsion systems. It turns out that there are intricacies and computational challenges for these systems, but work to prove this prediction is well underway and will be the topic of our forthcoming article.
 
\section*{Summary}
Relevant for microemulsions we have identified a truly non-classical critical point, named superflexibility critical point, which is linked to the ability of balanced L/L interfaces to become free of tension upon adsorption of surfactants. When the width of the interface exceeds the size of the surfactant the system cannot find the tensionless state and microemulsions cease to exist. This critical point (SFP) controls the phenomenology of microemulsion systems which excellently correlates with known microemulsion behavior \cite{gompper1995self,kahlweit1985phase}. (i) Sufficiently far from the critical point the Gaussian bending rigidity is negative and a lamellar microemulsion is expected. (ii) However, near the critical point the sign of the Gaussian bending rigidity changes. This means that we expect the sponge-like topology of the system. (iii) Importantly, when the Gaussian bending rigidity and the mean bending rigidity are of the same magnitude, the two moduli enter a scaling regime such that they vanish at the superflexibility critical point with scaling exponent unity. It is argued that the bicontinuous sponge phase is the most prominent marker for the superflexibility critical point (SFP). The experimental observation that the sponge phase is somewhat turbid \cite{clausse1981bicontinuous} is consistent with the system being critical. However, we do not expect classical critical scattering near the SFP, because, for example, the area expansion modulus $K_A$ does not vanish at the SFP.  

We are only at the beginning of understanding the many implications of the SFP, but it is foreseen that biomembrane rigidity and topology are under SFP control as well.

\matmethods{In this section we will briefly mention the characteristics of the SF-SCF framework, specify the model that is used and review the method that was employed to find the bending moduli. 
\subsubsection*{Methods}
Upon the implementation of the mean field approximation it becomes possible to write a closed expression for the mean field free energy (we use dimensionless units) for a molecularly inhomogeneous system \cite{leermakers2013bending,kik2010molecular,cosgrove1987configuration,hurter1993molecular,wijmans1992self,scheutjens1979statistical,fleer1993polymers}. This free energy $F$ is found \cite{scheutjens1979statistical} in terms of volume fraction $\varphi_x({\bf r})$ and complementary segment potential $u_x({\bf r})$ profiles for segment types $x= A,\ B$. Lagrange parameters $\alpha({\bf r})$ are introduced to implement the incompressibility of the system:
\begin{equation}
\begin{aligned}
F = - \ln Q([u]) - \sum_{x,{\bf r}}u_x(\textbf{r})\varphi_x (\textbf{r}) L(r) + F^{\rm int}([\varphi]) \\
+ \sum_{\textbf r}\alpha(\textbf{r}) \cdot\bigg{[}\sum_x L(r)\varphi_x(\textbf{r}) - L(r) \bigg{]}
\label{eq.scff}
\end{aligned}
\end{equation}
The system partition function which can be decomposed into single chain partition functions $q_i$ for molecule component $i = A_n,\ B_n,\ A_NB_N$: $Q=\Pi_i q_i^{n_i}/n_i!$. The molecular partition function $q_i$ can be computed in freely jointed chain (FJC) approximation from the (known) segment potentials [$u_x({\bf r})$], and $n_i$ is the number of molecules of type $i$ in the system. We use the Bragg Williams approximation, similarly as in regular solution theory, to find the interaction free energy  
\begin{equation}
F^{\rm int}=\chi\sum_r L(r) \varphi_A(r) \bigg{[} \varphi_B(r) + \frac{1}{6} \nabla^2 \varphi_B(r) -\varphi_B^b \bigg{]},
\end{equation} 
wherein $L(r)$ is the number of sites at coordinate $r$. The super index $b$ refers to the quantity in the bulk solution (which exists far from the interface). SCF solutions now involves optimizing the free energy ($F$) with respect to its variables, respectively segment potentials, volume fractions and Lagrange field. When $\partial F/\partial L(r) \varphi_x(\textbf{r}) = 0$, we find that the potentials must obey $u_x(\textbf{r})=\alpha(\textbf{r})+\partial F^{\rm int }/\partial L(r) \varphi_x(\textbf{r})$. Setting $\partial F/\partial u_x(\textbf{r}) =0$ shows the way to evaluate the densities: $L(r) \varphi_x(\textbf{r})=-\partial \ln Q/\partial u_x(\textbf{r})$. The propagator formalism\cite{fleer1993polymers}, which can be shown to do the same, is the preferred way to find these densities. Finally, $\partial F/\partial \alpha(\textbf{r}) =0$ says that the optimization should obey the compressibility relation $\sum_x \varphi_x(\textbf{r}) = 1$.

Numerical solutions that obey these requirements have the property that the potentials both determine and follow from the volume fractions profiles and vice versa and are said to be self-consistent. Besides structure of the interfaces (density distributions) we can evaluate the thermodynamic quantities. Importantly there is a closed equation for the grand potential density 
$\omega(r) = -\sum_i (\varphi_i(\textbf{r})-\varphi_i^b)/N_i - \alpha(\textbf{r}) -
 \chi(\varphi_A(\textbf{r})\big{[}\varphi_B(\textbf{r}) + \frac{1}{6} \nabla^2 \varphi_B(\textbf{r})\big{]}-\varphi_A^b \varphi_B^b)$, and the overall grand potential is found from $\Omega=\sum_r L(r) \omega(r)$. 
For the planar interface, the interfacial tension is found from $\gamma = \Omega/A_s =\sum_z \omega(z)$ with $A_s$ is the area of the L/L interface and $z$ the coordinate across the planar interface. 

In model calculations one can construct perfectly `symmetric' systems that have interfaces with no spontaneous curvature by construction. Of course in experiments such perfect systems are very hard to realise and we understand that experimental systems only can come close to these ideal systems by the addition of a suitable amount of a carefully selected cosolvent. We understand this complication and nevertheless elaborate the ideal system for reasons of clarity and simplicity. 

\subsubsection*{Model system}
As in our previous paper we consider two solvents A$_n$ and B$_n$ that have a solubility gap,  i.e. $\chi = \chi_{AB} >2/n$. Without mentioning otherwise we will use a small value $n=4$ for the solvent `molar volume'. We use symmetric block copolymer amphiphiles A$_N$B$_N$, so that there is only one interaction parameter in the system. The value of $N$ is chosen $N>n$ (default $N=20$) so that the surfactant is long enough to keep the solubility low in the bulk phases and the adsorption at the L/L interface is usually sufficiently strong (i.e. when $\chi \gg \chi^{\rm cr}$). This adsorption will lead to a decrease of the interfacial tension. As the system is completely symmetric the spontaneous curvature of the interface is zero. This means that the planar interface is the ground state. Spherical droplets will have a higher interfacial energy and these can only occur as a fluctuation (on the mean field level we can safely ignore the microemulsion droplet phases). We mention that the zero-tension interface is typical for the microemulsion state in which there is a large interfacial area. When, for example, the tension is positive, the system will respond by decreasing the interfacial area. This decrease will increase the number of surfactants per unit area, and the interfacial tension will decrease. Inversely, when the interfacial tension is negative, the response of the system is to increase the interfacial area. Then the number of surfactants per unit area will decrease. This leads to an increase of the tension. Only when the tension is zero, the system will neither increase nor decrease its area and the system is in equilibrium. We say that the tensionless interface is a `natural' attractor of the system, characteristic for the microemulsion state of the system. 

\subsubsection*{Estimation of bending rigidities}
For this model we reported in our previous paper an elegant route to unambiguously find the bending rigidities that feature in the Helfrich equation. Our basic idea was to use the scale invariance properties of the Helfrich equation to unravel the rigidities. We proved that this property allows the system to curve the interface homogeneously without affecting the chemical potentials in the system. Hence the interfacial tension is the characteristic function and the Helfrich Eqn reads
\begin{equation}
\gamma(J,K)=\frac{1}{2} \frac{\partial^2 \gamma}{\partial J^2} J^2 + \frac{\partial \gamma}{\partial K}K \equiv \frac{1}{2} \kappa  J^2 + \bar{\kappa}K
\label{Helfrich}
\end{equation}
Three geometries were found for which the scale invariance principle could be applied: 
(i) The first case is the spherical droplet with inner phase A$_n$- and outer phase the B$_n$ rich phase. We add the surfactant until the Laplace pressure vanishes. The total curvature energy is now given by $\Omega = 4 \pi R^2 \gamma(J,K)$. As the mean curvature in this case is $J=2/R$ and the Gaussian curvature $K=1/R^2$ we find that $\Omega= 4\pi (2\kappa + \bar{\kappa})$. Note that this result does not depend on $R$ (the hallmark for scale invariance).  

(ii) The second system is a case for which everywhere in the curved interface the mean curvature is zero. This occurs in a triple periodic bicontinuous Im3m cubic phase. The interface splits a unit cell exactly into two equivalent volumes, one filled by the A- and the other by the B-rich phase. We add surfactants until the chemical potentials are equal to the ones found for the planar tensionless interface. The curvature energy per unit cell is now only a function of the genus $g$ \cite{chern1944simple}, $\Omega = 4\pi (g-1) \bar{\kappa}$. For the Im3m phase we have $g=3$. Again, the size of the unit cell is not involved in this result and the curvature energy is scale invariant.

(iii) Interestingly there exist yet another system for which we can use scale invariance and this is for the {\it minimal} torus. A minimal torus is a torus for which the major radius $R_1$ and the inner radius $R_2$ have a fixed ratio $R_1/R_2=\sqrt{2}$.\cite{willmore1965note,marques2012min} Again we fill the torus inside with the A-rich phase while the B-rich phase is the exterior phase. As before we add surfactants until the Laplace pressure vanishes (cf. case i). The torus is an object with genus $g=1$ and thus has no contribution to the Gaussian curvature. The curvature energy for a minimal torus is given by $\Omega = 4\pi^2 \kappa$. 

All these cases can computationally be realised within the SCF framework and the results of this protocol gives consistent results for the bending rigidities. Moreover we have shown that the Gaussian bending modulus follows from the second moment over the grand potential density of the tensionless planar interface.

}

\acknow{We thank S. Safran for helpful discussions. This work is part of an Industrial Partnership Programme, `Shell/NWO Computational Sciences for Energy Research (CSER-16)', of the Foundation for Fundamental Research on Matter (FOM), which is part of the Netherlands Organisation for Scientific Research (NWO). Project number: 15CSER26. 

}
\showmatmethods{} 
\showacknow{} 

\bibliography{pnas-sample}

\begin{thebibliography}{1}

\bibitem{varadharajan2018sign}
Varadharajan R, Leermakers FAM (2018) Sign switch of gaussian bending modulus
  for microemulsions: A self-consistent field analysis exploring scale
  invariant curvature energies.
\newblock {\em Physical Review Letters} 120(2):028003.

\bibitem{chern1944simple}
Chern SS (1944) A simple intrinsic proof of the gauss-bonnet formula for closed
  riemannian manifolds.
\newblock {\em Annals of mathematics} pp. 747--752.

\bibitem{fenchel1940total}
Fenchel W (1940) On total curvatures of riemannian manifolds: I.
\newblock {\em Journal of the London Mathematical Society} 1(1):15--22.

\bibitem{safran1994statistical}
Safran SA (1994) {\em Statistical thermodynamics of surfaces, interfaces, and
  membranes}.
\newblock (Perseus Books) Vol.{}~90.

\bibitem{willmore1965note}
Willmore TJ (1965) Note on embedded surfaces.
\newblock {\em An. Sti. Univ. ``Al. I. Cuza'' Iasi Sect. I a Mat.(NS) B}
  11:493--496.

\bibitem{leermakers2013bending}
Leermakers FAM (2013) Bending rigidities of surfactant bilayers using
  self-consistent field theory.
\newblock {\em J. Chem. Phys.} 138(15):04B610.

\bibitem{oversteegen2000rigidity}
Oversteegen S, Blokhuis E (2000) Rigidity constants from mean-field models.
\newblock {\em The Journal of Chemical Physics} 112(6):2980--2986.

\bibitem{oversteegen2000thermodynamics}
Oversteegen SM, Leermakers FAM (2000) Thermodynamics and mechanics of bilayer
  membranes.
\newblock {\em Phys. Rev. E} 62(6):8453.

\end{thebibliography}


\begin{thebibliography}{10}

\bibitem{danielsson1981definition}
Danielsson I, Lindman B (1981) The definition of microemulsion.
\newblock {\em Colloids and Surfaces} 3(4):391--392.

\bibitem{yamamoto2001transparent}
Yamamoto J, Tanaka H (2001) Transparent nematic phase in a liquid-crystal-based
  microemulsion.
\newblock {\em Nature} 409(6818):321.

\bibitem{thiam2013biophysics}
Thiam AR, Farese~Jr RV, Walther TC (2013) The biophysics and cell biology of
  lipid droplets.
\newblock {\em Nature reviews Molecular cell biology} 14(12):775.

\bibitem{safran1986origin}
Safran S, Roux D, Cates M, Andelman D (1986) Origin of middle-phase
  microemulsions.
\newblock {\em Phys. Rev. Lett.} 57(4):491.

\bibitem{scriven1976equilibrium}
Scriven L (1976) Equilibrium bicontinuous structure.
\newblock {\em Nature (London)} 263(5573):123--125.

\bibitem{clausse1981bicontinuous}
Clausse M, Peyrelasse J, Heil J, Boned C, Lagourette B (1981) Bicontinuous
  structure zones in microemulsions.
\newblock {\em Nature} 293(5834):636--638.

\bibitem{dave2007self}
Dave H, et~al. (2007) Self-assembly in sugar--oil complex glasses.
\newblock {\em Nature materials} 6(4):287.

\bibitem{jones2012nanocasting}
Jones BH, Lodge TP (2012) Nanocasting nanoporous inorganic and organic
  materials from polymeric bicontinuous microemulsion templates.
\newblock {\em Polymer journal} 44(2):131.

\bibitem{varadharajan2018sign}
Varadharajan R, Leermakers FAM (2018) Sign switch of gaussian bending modulus
  for microemulsions: A self-consistent field analysis exploring scale
  invariant curvature energies.
\newblock {\em Physical Review Letters} 120(2):028003.

\bibitem{talmon1978statistical}
Talmon Y, Prager S (1978) Statistical thermodynamics of phase equilibria in
  microemulsions.
\newblock {\em The Journal of Chemical Physics} 69(7):2984--2991.

\bibitem{de1982microemulsions}
de~Gennes PG, Taupin C (1982) Microemulsions and the flexibility of oil/water
  interfaces.
\newblock {\em J. Phys. Chem.} 86(13):2294--2304.

\bibitem{huang1981critical}
Huang JS, Kim MW (1981) Critical behavior of a microemulsion.
\newblock {\em Physical Review Letters} 47(20):1462.

\bibitem{cazabat1982critical}
Cazabat A, Langevin D, Meunier J, Pouchelon A (1982) Critical behavior in
  microemulsions.
\newblock {\em Advances in Colloid and Interface Science} 16(1):175--199.

\bibitem{strey1990dilute}
Strey R, Schom{\"a}cker R, Roux D, Nallet F, Olsson U (1990) Dilute lamellar
  and l 3 phases in the binary water--c 12 e 5 system.
\newblock {\em Journal of the Chemical Society, Faraday Transactions}
  86(12):2253--2261.

\bibitem{szleifer1988curvature}
Szleifer I, Kramer D, Ben-Shaul A, Roux D, Gelbart WM (1988) Curvature
  elasticity of pure and mixed surfactant films.
\newblock {\em Phys. Rev. Lett.} 60(19):1966.

\bibitem{blokhuis1993van}
Blokhuis EM, Bedeaux D (1993) Van der waals theory of curved surfaces.
\newblock {\em Molecular Physics} 80(4):705--720.

\bibitem{gompper1992ginzburg}
Gompper G, Zschocke S (1992) Ginzburg-landau theory of oil-water-surfactant
  mixtures.
\newblock {\em Physical Review A} 46(8):4836.

\bibitem{gompper1995self}
Gompper G, Schick M, Milner S (1995) Self-assembling amphiphilic systems.

\bibitem{oversteegen2000rigidity}
Oversteegen S, Blokhuis E (2000) Rigidity constants from mean-field models.
\newblock {\em The Journal of Chemical Physics} 112(6):2980--2986.

\bibitem{supplemental}
(year?).
\newblock {\em See Supplemental Material http://link.aps.org/ which includes
  Refs. [27]}.

\bibitem{safran1999curvature}
Safran S (1999) Curvature elasticity of thin films.
\newblock {\em Advances in Physics} 48(4):395--448.

\bibitem{lekkerkerker1989contribution}
Lekkerkerker H (1989) Contribution of the electric double layer to the
  curvature elasticity of charged amphiphilic monolayers.
\newblock {\em Physica A: Statistical Mechanics and its Applications}
  159(3):319--328.

\bibitem{birshtein2008curvature}
Birshtein T, et~al. (2008) On the curvature energy of a thin membrane decorated
  by polymer brushes.
\newblock {\em Macromolecules} 41(2):478--488.

\bibitem{kahlweit1985phase}
Kahlweit M, Strey R (1985) Phase behavior of ternary systems of the type h2o
  oil nonionic amphiphile (microemulsions).
\newblock {\em Angewandte Chemie International Edition} 24(8):654--668.

\bibitem{kahlweit1988general}
Kahlweit M, et~al. (1988) General patterns of the phase behavior of mixtures of
  water, nonpolar solvents, amphiphiles, and electrolytes. 1.
\newblock {\em Langmuir} 4(3):499--511.

\bibitem{kahlweit1988properties}
Kahlweit M, Strey R, Haase D, Firman P (1988) Properties of the three-phase
  bodies in water-oil-nonionic amphiphile mixtures.
\newblock {\em Langmuir} 4(4):785--790.

\bibitem{kahlweit1986search}
Kahlweit M, Strey R, Firman P (1986) Search for tricritical points in ternary
  systems: water-oil-nonionic amphiphile.
\newblock {\em The Journal of Physical Chemistry} 90(4):671--677.

\bibitem{strey1996phase}
Strey R (1996) Phase behavior and interfacial curvature in water-oil-surfactant
  systems.
\newblock {\em Current Opinion in Colloid \& Interface Science} 1(3):402--410.

\bibitem{helfrich1985persistencelength}
Helfrich W (1985) Effect of thermal undulations on the rigidity of fluid
  membranes and interfaces.
\newblock {\em J. Phys. Paris} 46:1263--1269.

\bibitem{helfrich1987persistencelength}
Helfrich W (1987) Measures of integration in calculating the effective rigidity
  of fluid surfaces.
\newblock {\em J. Phys. Paris} 48:285--289.

\bibitem{cates1988random}
Cates M, Roux D, Andelman D, Milner S, Safran S (1988) Random surface model for
  the l3-phase of dilute surfactant solutions.
\newblock {\em EPL (Europhys. Lett.)} 5(8):733.

\bibitem{leermakers2013bending}
Leermakers FAM (2013) Bending rigidities of surfactant bilayers using
  self-consistent field theory.
\newblock {\em J. Chem. Phys.} 138(15):04B610.

\bibitem{kik2010molecular}
Kik RA, Leermakers FAM, Kleijn JM (2010) Molecular modeling of proteinlike
  inclusions in lipid bilayers: Lipid-mediated interactions.
\newblock {\em Phys. Rev. E} 81(2):021915.

\bibitem{cosgrove1987configuration}
Cosgrove T, Heath T, Van~Lent B, Leermakers F, Scheutjens J (1987)
  Configuration of terminally attached chains at the solid/solvent interface:
  self-consistent field theory and a monte carlo model.
\newblock {\em Macromolecules} 20(7):1692--1696.

\bibitem{hurter1993molecular}
Hurter PN, Scheutjens JM, Hatton TA (1993) Molecular modeling of micelle
  formation and solubilization in block copolymer micelles. 1. a
  self-consistent mean-field lattice theory.
\newblock {\em Macromolecules} 26(21):5592--5601.

\bibitem{wijmans1992self}
Wijmans C, Scheutjens J, Zhulina E (1992) Self-consistent field theories for
  polymer brushes: lattice calculations and an asymptotic analytical
  description.
\newblock {\em Macromolecules} 25(10):2657--2665.

\bibitem{scheutjens1979statistical}
Scheutjens JMHM, Fleer GJ (1979) Statistical theory of the adsorption of
  interacting chain molecules. 1. partition function, segment density
  distribution, and adsorption isotherms.
\newblock {\em J. phys. Chem} 83(12):1619--1635.

\bibitem{fleer1993polymers}
Fleer G, Stuart MAC, Scheutjens JMHM, Cosgrove T, Vincent B (1993) {\em
  Polymers at interfaces}.
\newblock (Springer Science \& Business Media).

\bibitem{chern1944simple}
Chern SS (1944) A simple intrinsic proof of the gauss-bonnet formula for closed
  riemannian manifolds.
\newblock {\em Annals of mathematics} pp. 747--752.

\bibitem{willmore1965note}
Willmore TJ (1965) Note on embedded surfaces.
\newblock {\em An. Sti. Univ. ``Al. I. Cuza'' Iasi Sect. I a Mat.(NS) B}
  11:493--496.

\bibitem{marques2012min}
Marques FC, Neves A (2014) Min-max theory and the willmore conjecture.
\newblock {\em Annals of Mathematics} 179:683--782.

\end{thebibliography}

\end{document}


\maketitle
\thispagestyle{firststyle}


\section{Mathematical representation of the interface}
To quantify the physics at the interfaces, we mathematically define the interface as an infinitely thin surface. Surfaces can be described in implicit form as $F(x,y,z)=0$. For such an implicit form one can obtain the normal to the surface by realizing that total derivative of $F$ vanishes: $dF = dr.\nabla F =0$, where $dr$ is a vector connecting two points on the surface. As $dr$ is a tangential vector, $\nabla F$ must be an orthogonal vector and points in the direction of normal to the surface. Thus, unit normal to the surface can be given by
\begin{equation}
\hat{n} = \nabla F/|\nabla F|
\label{eq1}
\end{equation}
Reimann defined curvature of the surface as the change in the normal vector as one moves along the surface. Such a change is given by the curvature tensor $Q$. If we move along the surface by a distance $dr$, the normal $\hat{n}$ changes by an amount $d\hat{n} = dr. Q$ where elements of tensor $Q$ is obtained by differentiating equation \ref{eq1}.

\begin{figure}[h]
\centering
\begin{tikzpicture}
  \begin{axis}[
  	xlabel={x},
    ylabel={y},
    zlabel={z}, width=0.25\textwidth
    ]
    \addplot3[surf]{x^2-y^2};
    \draw[->,black,thick] (axis cs:0,0,0) -- node[black,right]{$\hat{n}$} (axis cs:0,0,15);
    \draw[->,black,thick] (axis cs:-3,-5,0) -- node[black,below]{$\hat{n}$} (axis cs:-4,-6,0.05);
    \node at (axis cs:0,0,25){$F(x,y,z)=0$};
  \end{axis}
\end{tikzpicture}
\caption{Illustration of mathematically defined interface $F(x,y,z)=0$, with normal vectors $\hat{n}$ shown for two sample points. }
\end{figure}
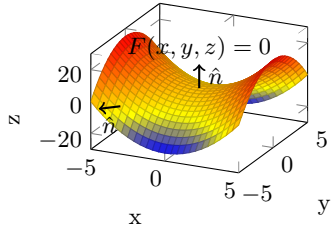

Curvature tensor $Q_{ij}$ is thus given by,
\begin{equation}
Q_{ij} = \frac{1}{|\nabla F|}\bigg{[}F_{ij} - \frac{F_i|\nabla F|_j}{|\nabla F|}\bigg{]}
\end{equation}
where $F_i = \partial F/\partial r_i$. Such three dimensional tensors have three invariants under similarity transformations: trace, sum of principal minors and determinant. We know from explicit calculations that one of the eigen values is zero and thus the determinant. Thus, curvature tensor is a singular matrix. Two of the eigen values of $Q$ have dimensions of inverse length and are defined as the principal curvatures,  namely $1/R_1, 1/R_2$. This allows us to define the invariant mean curvature as the trace of $Q$: $H = 1/2(1/R_1 +1/R_2)$. In our expressions we $J = 2H$ as mean curvature. This adoption is purely notational and will not affect the physical results. Similarly, second invariant of $Q$ is defined as Gaussian curvature: $K = 1/(R_1R_2).$

\section{Calculation of the interfacial width}
There are several ways to define interfacial width. To avoid ambiguity we explain the way we have defined and calculated the interfacial width in our results. It should be noted, however, the physics remains generic irrespective of the way the interfacial width is defined. In our results, interfacial width $W$ is calculated from the volume fraction profiles of the bulk phases(either oil or water) as follows,
\begin{equation}
W = \frac{\Delta \varphi}{\varphi_{z=\frac{-1}{2}}-\varphi_{z=\frac{1}{2}}}
\label{eq3}
\end{equation}
\begin{figure}[h]
\centering
\begin{subfigure}{0.45\textwidth}
\centering
\begin{tikzpicture}
\begin{axis}[
  x unit={},
  y unit={},
  xlabel={z},
  ylabel={$\varphi$}, 
  y label style={at={(-0.275,0.5)},anchor=south},
  legend pos=north west,
  minor x tick num=1,
  minor y tick num=1, xmin=-30, xmax=30, ymin=0, ymax=1, width=0.5\textwidth]
\fill [fill=yellow, opacity=0.5] (axis cs: 5,0) rectangle (axis cs: -5,1);
\addplot[color=red, thick] table [x index=0,y index=1] {phi_fc.dat};
\addplot[color=blue, thick] table [x index=0,y index=2] {phi_fc.dat};
\addplot[color=magenta, thick] table [x index=0, y index=3] {phi_fc.dat};
\node at (axis cs: 30, 0.65) [anchor=north east] {(a)};
\draw[|<->|,black,thick] (axis cs:5,0.4) -- node[above]{$W$} (axis cs:-5,0.4);
\end{axis}
\end{tikzpicture}
\end{subfigure}
\begin{subfigure}{0.35\textwidth}
\centering
\includegraphics[width=0.45\linewidth]{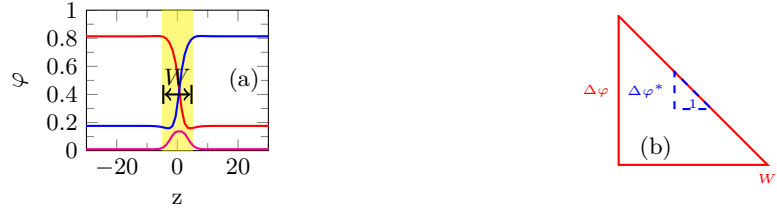}
\end{subfigure}
\caption{(a) Volume fraction profiles of oil (orange) water(blue) and surfactants(blue) far from superflexibility point ($\chi = 0.6$). (b) Similar triangles representing the width as unknown, similarity rule provides us with the equation \ref{eq3}.}

\end{figure}
It should be noted that volume fraction profiles of oil and water phase are symmetric, thus using either of them to compute the interfacial width should give us similar values. However, interfacial width computed from any other profile, ex. surfactant volume fraction, grand potential etc., when used to define the interfacial width will provide a quantitatively a different picture and yet remains qualitatively the same. 

\section{Determination of bending rigidities}
For interested reader we refer our earlier article \cite{varadharajan2018sign}, in which we had published our results on scale invariant surfaces to unambiguously determine the bending rigidities of surfacted liquid liquid interfaces. In this section, we will provide a brief overview of the procedure and the essential details involved in direct estimation of the gaussain bending modulus from Im3m cubic phase. 

\subsection{Computational methodology}
For a tensionless interface ($\gamma(0,0) = 0$) in saddle shape, with zero mean curvature ($J = 0$) and spontaneous curvature($J_0 = 0$), the energy of the interface (single Reimann manifold) from Helfrich's equation, can be shown as
\begin{equation}
\iint \gamma_G(J,K) dA = \bar{k} \iint K dA
\label{eq.energyim3m}
\end{equation}
Now we use Gauss-Bonnet theorem for compact, boundary-less Riemann manifold. 
Theorem states integral of curvature over area of the manifold is equal to $2\pi$ times the Euler characteristic of the manifold\cite{chern1944simple,fenchel1940total}
\begin{equation}
\iint K dA = 4\pi (1-g)
\label{eq.gaussbonnet}
\end{equation}
where g is the genus of the manifold. 
For Im3m cubic phase the genus (g) is $3$.
Thus for an interface in saddle shape, interfacial energy is $-8\pi$ times the Gaussian bending modulus ($\bar k$)
\begin{equation}
\Omega_{Im3m}^s = - 8\pi \bar{k}
\label{eq.gaussbonnetim}
\end{equation}
It is clear from equation \ref{eq.gaussbonnetim} that interface with positive Gaussian bending modulus ($\bar k > 0$) prefers to form handles and complicated shapes, consistent with findings reported in literature \cite{safran1994statistical}.

To numerically generate such saddle shape in mean field approach, a special strategy is required.We imposed special boundary conditions that directs the system towards the saddle configuration, which is used as an initial guess for the calculations. As soon as initial guess is generated, Symmetry or Mirror boundary condition is forced on $6$ faces of 3D-Box or elementary box. Since by generating $1/8$ of the unit cell, we ensure using symmetry boundary conditions, the effects of entire unit cell is captured. All calculations are performed in double precision so that the thermodynamical quantities obtained are not affected by round-off errors. It should also be noted since we only calculate $(1/8)^{th}$ of the manifold, the energy of the interface should be normalized by a factor of $8$.
\begin{equation}
\big{[}\Omega_{Im3m}^s\big{]}_{3D-Box} = -\pi \bar{k}
\label{eq.gaussbonnetimelem}
\end{equation}
It could be observed that the saddle shaped interface, inferred as two dimensional Riemann manifold without boundary, crosses all the faces.
These crossings are signature of handle formation in the system, unlike planar interface where the interface or manifold crosses only $4$ faces.
Since we are trying to represent a complex saddle shaped manifold in a Cartesian lattice, lattice artifacts become unavoidable in our calculations.
Due care has been taken to reduce the lattice artifacts, by optimizing the system size.
A quantitative measure of tendency of any surface to deviate from spherical shape is provided by Willmore energy ($W$)
\begin{equation}
W = \frac{1}{4} \iint J^2dA - \iint KdA
\label{eq.willmoreenergy}
\end{equation}
Willmore energy \cite{willmore1965note}, of an ideal Im3m cubic phase, cf. equation \ref{eq.willmoreenergy}, contains only Gaussian curvature energy, as mean curvature is zero ($J = 0$). 
In calculations, representation of Im3m phase in a discretized Euclidean space gives rise to contribution in Willmore energy ($W$) from residual mean curvature ($J \approx 0$).
With increase in box size, the residual mean curvature energy increases whereas the surface in discretized space is better represented and vice versa.
Hence the box size in calculations are optimized to ensure that surface is better represented with residual mean curvature energy remains negligible. 
From previous experiments and experience \cite{leermakers2013bending}, the optimal value of box size for a surfactant chain-length ($N=30$) is found to be $m_x = m_y = m_z = m = 50$.

\section{Evolution of pressure profiles}
In Fig. \ref{sf3}, the evolution of pressure profile on approaching towards SFP is shown. As the SFP is approached the peak in the pressure profile decreases. It should also be noted that the magnitude of y$-$axis decreases drastically as the adsorbed amount of surfactant decreases. The widening of the dip in  $\omega(z)$ eventually prevents the system from reaching a tensionless state. To understand the effect of concentration, it is possible to decrease computational box size. To understand such asymmetry in interfaces will be one of our potential future works.

\begin{figure}[t]
\centering
\begin{tikzpicture}
\begin{axis}[
  x unit={\it b},
  y unit={},
  xlabel={z},
  ylabel={$\omega$}, 
  y label style={at={(-0.275,0.5)},anchor=south},
  legend pos=north west,
  minor x tick num=1,
  minor y tick num=1, xmin=-30, xmax=30, width=0.25\textwidth]
\addplot[color=red, thick] table [x index=0,y index=1] {pp7.dat};
\node at (axis cs: 30, -0.000001) [anchor=north east] {(a)};
\end{axis}
\end{tikzpicture}
\begin{tikzpicture}
\begin{axis}[
  x unit={\it b},
  y unit={},
  xlabel={z},
  ylabel={$\omega$}, 
  y label style={at={(-0.275,0.5)},anchor=south},
  legend pos=north west,
  minor x tick num=1,
  minor y tick num=1, xmin=-30, xmax=30, width=0.25\textwidth]
\addplot[color=red, thick] table [x index=0,y index=1] {pp8.dat};
\node at (axis cs: 30, -0.000001) [anchor=north east] {(b)};
\end{axis}
\end{tikzpicture}

\begin{tikzpicture}
\begin{axis}[
  x unit={\it b},
  y unit={},
  xlabel={z},
  ylabel={$\omega$}, 
  y label style={at={(-0.275,0.5)},anchor=south},
  legend pos=north west,
  minor x tick num=1,
  minor y tick num=1, xmin=-30, xmax=30, width=0.25\textwidth]
\addplot[color=red, thick] table [x index=0,y index=1] {pp9.dat};
\node at (axis cs: 30, -0.000001) [anchor=north east] {(c)};
\end{axis}
\end{tikzpicture}
\begin{tikzpicture}
\begin{axis}[
  x unit={\it b},
  y unit={},
  xlabel={z},
  ylabel={$\omega$}, 
  y label style={at={(-0.275,0.5)},anchor=south},
  legend pos=north west,
  minor x tick num=1,
  minor y tick num=1, xmin=-30, xmax=30, width=0.25\textwidth]
\addplot[color=red, thick] table [x index=0,y index=1] {pp10.dat};
\node at (axis cs: 30, -0.000001) [anchor=north east] {(d)};
\end{axis}
\end{tikzpicture}

 \caption{Evolution of Pressure profiles as the Flory-huggins interaction parameter is varied to approach the Superflexibility critical point. Results for $\chi$ values of $0.55$,$0.54$,$0.53$,$0.52$ in subplots (a),(b),(c),(d) respectively. 
 }
 \label{sf3}
\end{figure}
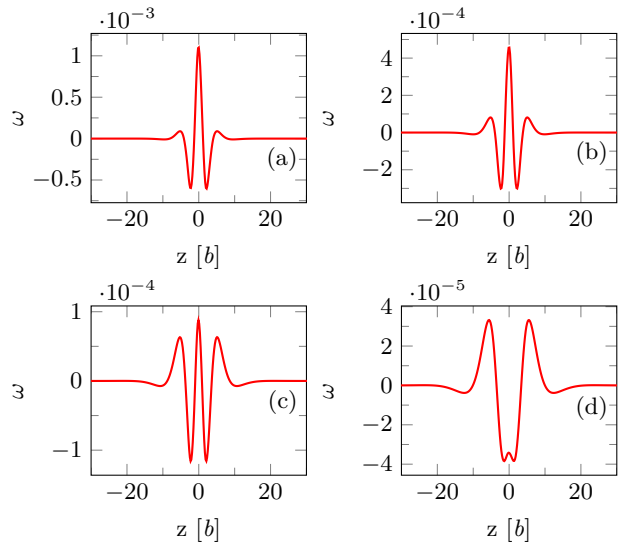

\begin{figure}[t]
\centering
\begin{tikzpicture}
\begin{loglogaxis}[
  x unit={},
  y unit={\it k_BT},
  xlabel={$\chi - \chi^{\rm cr}$},
  ylabel={$\gamma$}, 
  xmax=1e-1,
  xmin=1e-3,
  legend pos=north west,
  minor x tick num=1,
  minor y tick num=1, width=0.25\textwidth]
\addplot[color=black, thick] table [x index=0,y index=2] {llscaling.dat};
\draw[gray,thick] (axis cs:1e-3,1e-5) -- node[black,right]{$\frac{dy}{dx}=3/2$} (axis cs:1e-1,1e-2);
\node at (axis cs: 1e-3, 2e-2) [anchor=north west] {(a)};
\end{loglogaxis}
\end{tikzpicture}
\begin{tikzpicture}
\begin{loglogaxis}[
  x unit={},
  y unit={\it k_BT},
  xlabel={$\chi - \chi^{\rm cr}$},
  ylabel={$\kappa$, $\bar{\kappa}$}, 
  xmax=1e-1,
  xmin=1e-3,
  legend pos=north west,
  minor x tick num=1,
  minor y tick num=1, width=0.25\textwidth]
\addplot[color=black, thick] table [x index=0,y index=1] {llscaling.dat};
\addplot[color=black, thick, densely dashdotted] table [x index=0,y index=3] {llscaling.dat};
\draw[gray,thick] (axis cs:1e-3,5e-3) -- node[black,right]{$\frac{dy}{dx}=1/2$} (axis cs:1e-1,5e-2);
\node at (axis cs: 1e-3, 1e-1) [anchor=north west] {(b)};
\end{loglogaxis}
\end{tikzpicture}

 \caption{Extended results of liquid-liquid interface without surfactants. Bulk chain length is 4. (a) Scaling of interfacial tension close to bulk critical point. Slope is indicated in gray line. Numerical calculations follow van der Waals scaling, $\gamma = 1/\sqrt{6} (\chi-\chi^{\rm cr})^{3/2}$ (b) Scaling of bending rigidities, $\kappa$ (solid) and $\bar{\kappa}$ (dash dotted). Slopes is indicated in gray line. Results follow van der Waals theory $\kappa$, $\bar{\kappa} \approx (\chi-\chi^{\rm cr})^{1/2}$ }
 \label{sf4}
\end{figure}
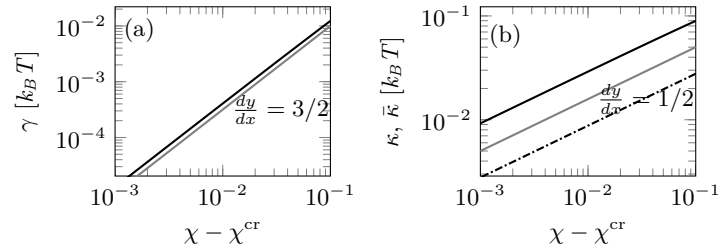

\section{Liquid-Liquid interface}

Behavior of the liquid-liquid interfaces near critical regime was studied  and published earlier by our group \cite{oversteegen2000rigidity,oversteegen2000thermodynamics}. Here we show extended results of the earlier work, and comparison with van der Waals equation. It should be noted that since there is no constraint in interfacial widening in non-surfacted interfaces, the interface exists till the bulk critical point is reached. Thus we were able to calculate the bending rigidities very close to the bulk critical point as shown in the Fig.\ref{sf4}.

\bibliography{pnas-sample_supp}